\documentclass[prl,twocolumn,showpacs,floatfix,superscriptaddress]{revtex4}
\usepackage{times}
\usepackage{graphicx}
\usepackage{amsmath, amssymb}

\begin{document}
\title{Aluminene as Highly Hole Doped Graphene}

\author{C. Kamal}
\author{Aparna Chakrabarti}
\affiliation{Indus Synchrotrons Utilization Division, Raja Ramanna Centre for Advanced
Technology, Indore 452013, India}
\author{Motohiko Ezawa}
\affiliation{Department of Applied Physics, University of Tokyo, Hongo 7-3-1, 113-8656, Japan}

\begin{abstract}
Monolayer structures made up  of purely one kind of atoms are fascinating. 
Many kinds of honeycomb systems including carbon, silicon, germanium, tin, phosphorus and arsenic have been shown to be stable. However, so far the structures are restricted to group IV and V elements. In this letter, we systematically investigate the stability of monolayer structures made up of aluminium, in four different geometric configurations (planar, buckled,  puckered and triangular), by employing density functional theory based electronic structure calculation. Our results on cohesive energy and phonon dispersion predict that only planar honeycomb structure made up of aluminium is stable. We call it "aluminene" according to the standard naming convention. It is a metal. Results of electronic band structure  suggest that it may be regarded as a highly hole doped graphene. We also present the tight-binding model and the Dirac theory to discuss the electronic properties of aluminene.
\end{abstract}


\maketitle

Monolayer structures made up of purely one kind of atoms possess many novel properties different from their bulk counterparts. The best example is graphene, which is a honeycomb monolayer of carbon\cite{NetoRev,KatsText}. The other group IV elements such as silicon, germanium and tin also form a honeycomb monolayer known as silicene, germanene and stanene, respectively\cite{LiuPRB,LiuPRL,EzawaPRL,stanene}.
Their geometric structures are buckled, where we can control the band gap by applying transverse electric field\cite{ni, Falko,EzawaNJP,Kamal}. 
They are expected to be topological insulators\cite{LiuPRL,EzawaPRL}. Very recently, silicene is experimentally demonstrated to act as a field-effect transistor at room temperature\cite{SiliceneFET}. Phosphorene,  honeycomb monolayer of phosphorus atoms, has experimentally been synthesized by exfoliating black phosphorus\cite{Li,Liu,Xia}. Its geometric structure is  puckered.  It is a direct-band gap semiconductor and possesses high hole mobility\cite{qiao}. One can also tune the band gap by increasing the number of layers\cite{qiao,Tran,Rudenko}. Another group V  honeycomb monolayer system (buckled and puckered structures) made up of arsenic atoms, called arsenene, has recently been predicted to be stable\cite{Arsenene}. 

It is note worthy that, so far there is no report available on monolayer structure composed of purely group III elements. Though there exists a study on  boron based finite triangular structure with a hexagonal hole,  it lacks translational invariant symmetry to form a periodic system\cite{Borophene}. Thus, it is important to probe whether group III atom based monolayer systems are stable and then investigate the physical properties of the stable ones. 

With this motivation, we investigate the stability of monolayer structures made up of group III elements (B, Al, Ga and In) and then study the geometric, electronic and vibrational properties by employing density functional theory (DFT)\cite{DFT} based calculations. We consider four different geometrical configurations, namely, (a) planar, (b) buckled,  (c) puckered and (d) triangular geometries. Our DFT based  phonon and cohesive energy calculations predict that among the above mentioned four configurations of group III elements, only the planar honeycomb monolayer made up of aluminium, as that of graphene, is stable. 
We call it "aluminene" in analogous to graphene.  We also report the geometric and electronic properties of aluminene with alkali metal (M = Li and Na) adsorption. Further, we construct an 8-band tight-binding model and the Dirac theory to discuss the electronic properties of aluminene. 

We use QUANTUM ESPRESSO package\cite{QE} for performing a fully self-consistent DFT calculations\cite{DFT} by solving the standard Kohn-Sham equations.  For exchange-correlation (XC) potential,  the generalized gradient approximation given by Perdew-Burke-Ernzerhof\cite{PBE} has been utilized. We use Rappe-Rabe-Kaxiras-Joannopoulos  ultrasoft pseudopotentials\cite{rrkj} for Al, Li and Na atoms which include the scalar-relativistic effect\cite{QE-lib}. Kinetic energy cutoff of 50 Ry  has been used for electronic wave functions. We adopt Monkhorst-Pack scheme for \textbf{k}-point sampling of Brillouin zone integrations with 61 $\times$ 61 $\times$ 1, 51 $\times$ 51 $\times$ 1  and 51 $\times$ 41 $\times$ 1 for the triangular / buckled, planar and puckered systems, respectively. The convergence criteria for energy in SCF cycles is chosen to be 10$^{-10}$ Ry. The geometric structures are fully optimized by minimizing the forces on individual atoms with the criterion that the total force on each atom is below  10$^{-3}$ Ry/Bohr.  In order to mimic the two-dimensional system, we employ a super cell geometry with a vacuum of about 18 \AA{}  in the direction perpendicular to the plane of monolayers so that the interaction between two adjacent unit cells in the periodic arrangement is negligible. To calculate the phonon spectra, we employ the density functional perturbation theory (DFPT)\cite{dfpt} implemented in PHonon code of  the QUANTUM ESPRESSO package. The dynamical matrix is estimated with a  7 $\times$ 7 $\times$ 1 mesh of \textbf{Q}-points in the Brillouin zone.

 \begin{figure*}[]
 \begin{center}
 \includegraphics[width=0.95\textwidth]{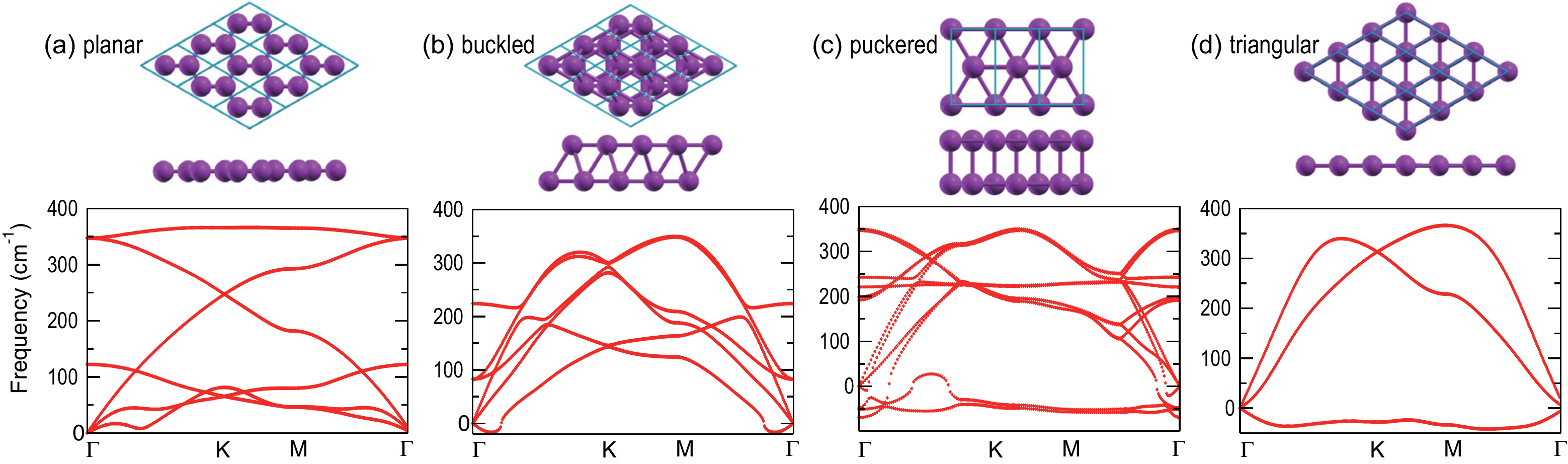}
 \end{center}
 \caption{Optimized geometries (top and side views) and phonon dispersion of aluminene in   (a) planar, (b) buckled, (c) puckered and (d) triangular configurations.  Planar aluminene is globally stable since the global minimum exists at the $\Gamma$ point. There are negative phonons (corresponding to transverse acoustic modes) except for the planar configuration, indicating their instability. }
 \end{figure*}

\textbf{Stability:} In order to study the stability of aluminene, we have carried out phonon dispersion calculations for all the geometric configurations. The fully optimized geometric structures and their corresponding phonon dispersion spectra are given in Fig.1. Our calculations predict that planar aluminene with space group $P6/mmm$ forms a stable structure since it contains only positive frequencies for all the vibrational modes [Fig.1(a)]. On the other hand, the remaining geometric configurations possess structural instabilities in the transverse acoustic modes due to the presence of imaginary frequencies (shown as negative values). Cohesive energy for planar aluminene is estimated to be -1.956 eV and the negative value indicates that this is a bound system. Our calculations yield 4.486 \AA{} and 2.590 \AA{} for lattice constant and Al-Al bond length for the planar structure, respectively (see Table I).  After the geometry optimization the buckled (puckered) structure has become AB (AA) bilayer of triangular lattice  [Fig.1(b) and (c)]. The relative displacement between A and B layers is along the (1/3, 2/3, 0) direction.  Since the aluminene in triangular, buckled and puckered configurations are not stable, as observed  from their phonon spectra, we do not consider them for further analysis. 

We have also carried out phonon dispersion calculations for the planar, buckled, puckered and triangular configurations made up of other group III elements (boron, gallium and indium). Our results show that there are no stable configurations.

\begin{table}[t]
\begin{center}
\caption{The results for optimized geometries of aluminene and with alkali metals (M = Li and Na) are given below.}
 \begin{tabular}{lccccccc}
\hline
\hline
Structure	 &	  Cohesive&Lattice  & \multicolumn{2}{c}{Bond length (\AA{})} \\  \cline{4-6}
			&energy  &constant 	& Al-Al & Al-M & M-M\\
		&(eV/atom)& $a$ (\AA{})	&  & & \\
\hline
Aluminene 		&-1.956	&4.486	&	2.590	&	-	&	-	\\
Aluminene + Li	&-2.151	&4.517	&	2.608	&	2.783	&	-	\\
Aluminene + Na	&-1.899	&4.531	&	2.616	&	3.197	&	-	\\
Aluminene + 2 Li	&-2.218	&4.420	&	2.552	&	2.908	&	2.790	\\
Aluminene + 2 Na&-1.846	&4.554	&	2.629	&	3.217	&	3.709	\\
\hline
\hline
\end{tabular}
\end{center}
\end{table}

\textbf{Electronic structure:}
We present the results of the electronic band structure, density of states (DOS) and partial DOS (PDOS) for planar aluminene in Fig.2(a). Aluminene behaves as a metallic system due to the partial occupancies in the $\sigma$ as well as $\pi$ bands.
We wish to note that the other honeycomb monolayer materials studied so far are either semimetal or semiconductor. At the high symmetric $K$ point in Brillouin zone, two Dirac cones occur at energies $1.618$ eV and $-4.274$ eV. The former is due to the $\pi$ bonds purely made up of the $p_z$ orbital and the latter is due to $\sigma$ bonds which has strong contribution from the $s$ orbital. At the Fermi energy, there exist two $\sigma$ bands and we call them the $\sigma_1$ and $\sigma_2$ bands as shown in Fig.3(a).  The $\sigma_2$ band is purely made up of $p_x$ and $p_y$ orbitals, while the mixing of the $s$ orbital is not negligible in the $\sigma_1$ band.

The electronic band structure of aluminene closely resembles that of graphene\cite{Saito}, but the locations of Dirac points in these two systems are different. In the case of graphene, the Dirac point lies exactly at the Fermi level where the bands due to the bonding  orbitals ($\sigma$ and $\pi$) are completely filled and those of anti-bonding orbitals are completely empty ($\sigma^*$ and $\pi^*$). On the other hand, in aluminene, the bands due to the bonding orbitals ($\sigma$ and $\pi$) are not completely filled and thus the Dirac point lies about 1.618 eV above the Fermi level.  This is due to the fact that the number of valence electrons in Al (trivalent) is smaller than that of C (tetravalent). As a result, aluminene can be regarded as a highly hole doped graphene.   However, the important difference is that the Fermi energy is highly shifted in aluminene.  Here, we wish to point out a remarkable feature of aluminene with respect to van-Hove singularities.   The van-Hove singularity of the $\sigma$ band exists very close to the Fermi energy at  $E = -0.212$ eV.  It may be possible to tune the Fermi energy around the van-Hove singularity and to make the electrical conductivity very large by applying gate voltage.

 \begin{figure}[t!]
 \begin{center}
 \includegraphics[width=0.4\textwidth]{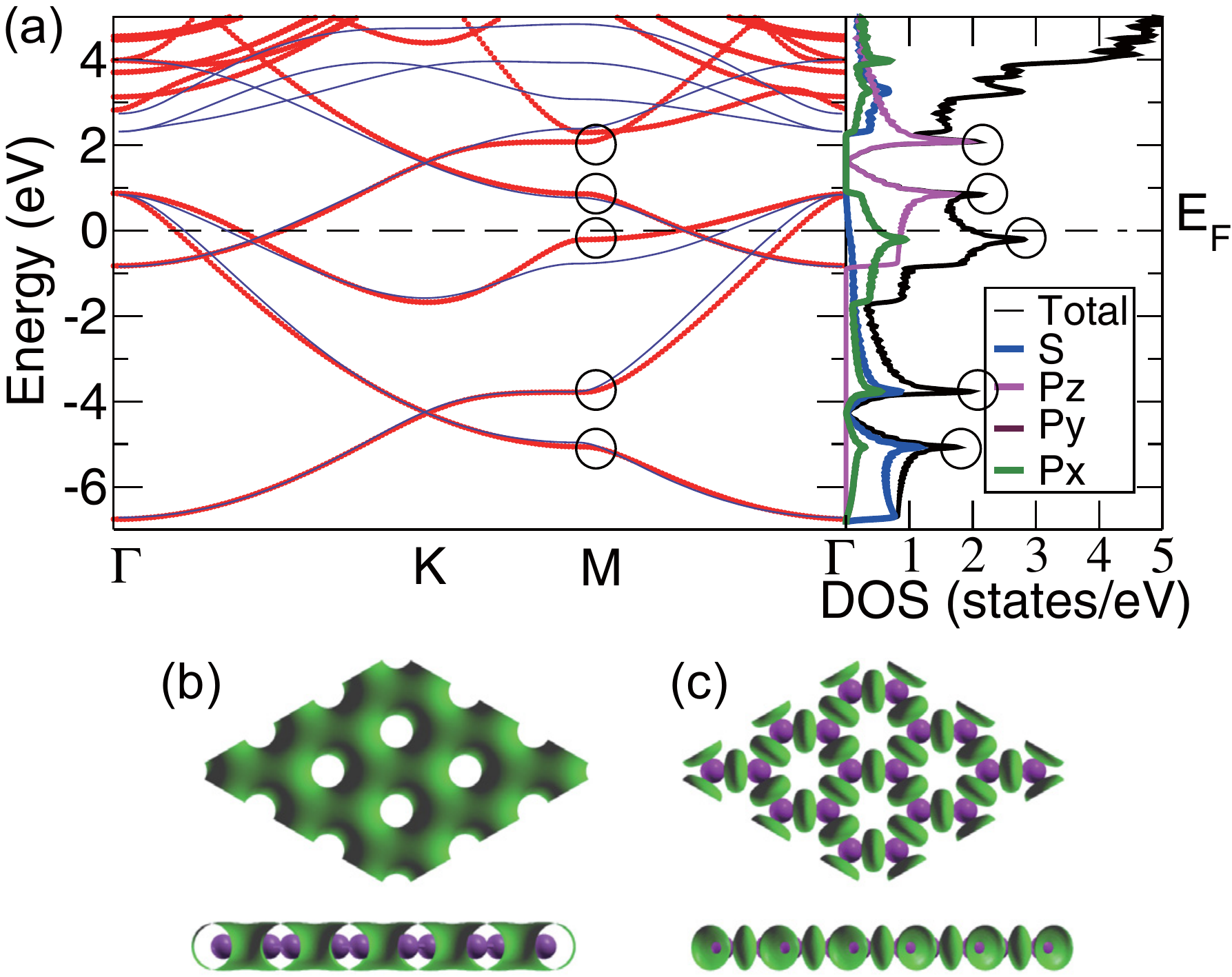}
 \end{center}
 \caption{Planar aluminene. (a) Left panel: electronic band structure; right panel: total and partial DOS. 
 Red curves are obtained from the first-principles calculation.
 Thin black curves are obtained from the tight-binding model. 
 Circles represent van-Hove singularities, where the DOS is divergent. 
 (b) Valence charge density. (c) Difference between the charge density of system and their atomic charge densities (top and side views).  }
 \end{figure}

To understand the nature of bonding in aluminene, we have calculated the valence charge density [Fig.2(b)]. Except the hollow regions of hexagon, the charge is nearly uniformly extended over the aluminene plane which contributes to the conductivity.  
We have also calculated the difference between the charge densities of system and its atomic constituents [Fig.2(c)]. It clearly indicates the presence of covalent bonds between Al atoms in aluminene. 

The transport properties of a metallic system are governed by the Fermi surface.
The calculated Fermi surface for aluminene is given in Fig.3(b). It consists of three closed loops, corresponding to the $\pi$, $\sigma_1$ and $\sigma_2$ bands. The momentum vectors at which the bands cross the Fermi level along high symmetric $k$-points are shown in Fig.3(a). Magnitudes of momentum vectors along the $\Gamma$-K and $\Gamma$-M directions for the $\sigma_1$ and $\pi$ bands are  equal ($k_1 = k_2$ and $k_5 = k_6$). Furthermore,    Fermi surfaces for these two bands are circular in shape around the $\Gamma$ point.  Thus, the electrons in these bands behave as free-particles. However, the magnitudes of momentum vectors along the $\Gamma$-K and $\Gamma$-M directions for the $\sigma_2$ band are not equal ($k_3 \ne k_4$). The Fermi surface for this band is highly hexagonal warped.  We comment on these Fermi surfaces further when we analyze the Dirac theory, given below.

 \begin{figure}[t!]
 \begin{center}
 \includegraphics[width=0.5\textwidth]{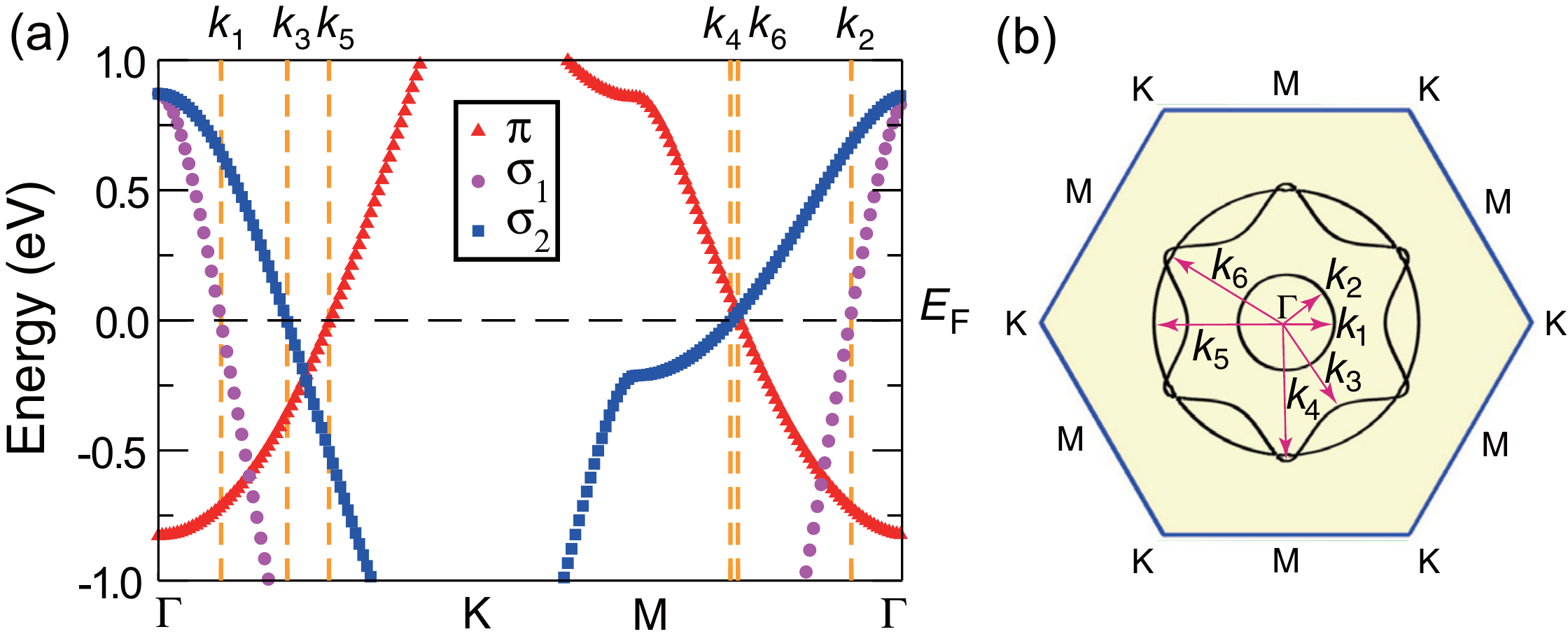}
 \end{center}
 \caption{Planar aluminene. (a) Electronic bands which cross the Fermi level and (b) their Fermi surfaces (curves in 2D). One Fermi surface is hexagonal warped, while the other two Fermi surfaces are of circular shape.}
 \end{figure}

\textbf{Tight-binding model:}
Electron configuration of aluminium is [Ne] 3s$^2$3p$^1$. 
We may construct an 8-band tight-binding model including the 4 orbitals:
$3s$, $3p_x$, $3p_y$ and $3p_z$.
The tight-binding Hamiltonian is given by
\begin{equation}
H=\sum_{ij}\sum_{\alpha\beta}t^{\alpha\beta}_{ij}c^{\dagger}_{i\alpha}c_{j\beta},
\end{equation}
where $t^{\alpha\beta}_{ij}$ denotes the nearest-neighbor transfer integral between the $\alpha$ orbital at the site $i$ and the $\beta$ orbital at the site $j$.
We have derived the  six independent Slater-Koster parameters ($\varepsilon_p$,   $\varepsilon_s$, $V_{pp\pi}$, $V_{pp\sigma}$, $V_{ss\sigma}$ and $V_{sp\sigma}$) by fitting the DFT based band structure at the $\Gamma$ and $K$ points, and present them in Table II.
The tight-binding model reproduces the global band structure well, but not precise Fermi momenta, as has been shown in Fig.2(a).

\begin{table}[t]
\begin{center}
\caption{The Slater-Koster parameters are obtained by fitting the DFT based electronic band structure at the $\Gamma$ and $K$ points.}
 \begin{tabular}{lccccccc}
 \hline
\hline
Structure	&	$\varepsilon_p$&  $\varepsilon_s$&$V_{pp\pi}$& $V_{pp\sigma}$ & $V_{ss\sigma}$ &$V_{sp\sigma}$\\  
\hline
Aluminene		&1.618	&-1.969	&-0.815	&-1.313	&-1.596	&-1.737\\
Aluminene + Li	&1.040	&-2.490	&-0.796	&-1.181	&-1.535	&-1.693 \\
Aluminene + Na	&0.862	&-2.256	&-0.854	&-0.985	&-1.569	&-1.620 \\
Aluminene + 2 Li	&-0.128	&-3.029	&-0.750	 &-1.017& -1.601	&-1.582 \\
Aluminene + 2 Na	&0.089	&-3.184	&-0.867	&-1.013	&-1.395	&-1.347	\\
Graphene\cite{Fabian}	&0	&-8.370	&-3.070		&-6.050	&-5.729	&-5.618	\\
\hline
\hline
\end{tabular}
\end{center}
\end{table}

\textbf{Dirac theory:} 
We proceed to construct the Dirac theory to describe the band structure near the Fermi energy.
First, the Fermi surface of the $\sigma_2$ band is highly anisotropic [see Fig.3(b)].
Hexagonal warping occurs due to the hexagonal symmetry at the $\Gamma$ point. 
It is well described by the hexagonal warped Dirac Hamiltonian, 
which was originally proposed to describe the Dirac fermion on the surface of the 3D topological insulator\cite{Hexa},
\begin{equation}
H=\hbar v(k_{x}\sigma _{y}-k_{y}\sigma _{x})+\gamma (k_{+}^{3}+k_{-}^{3})\sigma
_{z}+\varepsilon _{p},
\label{hexa}
\end{equation}
where 
$v=0.678a/\hbar$ is the velocity, 
$\gamma =0.237a^3$ describes the hexagonal warping
effects, $\varepsilon _{p}$ is the energy shift, $k_{\pm}=k_x\pm ik_y$ and $\mathbf{\sigma }$ denotes the Pauli matrix. 
The energy is given by 
\begin{equation}
E=\varepsilon _{p}\pm \sqrt{(\hbar vk)^{2}+(\gamma(k_{x}^{3}-3k_xk_{y}^{2}))^{2}}.
\end{equation}
It explains well the hexagonal warped Fermi surface shown in Fig.3(b), where the values of 
the Fermi momenta $k_3=0.269\times (2\pi /a)$ along the $K$ direction and $k_4=0.380\times (2\pi /a)$ along the $M$ direction are  different ($k_4>k_3$).

Next, 
the $\sigma_1$ band yields a circular shaped Fermi surface with $k_1=k_2=0.129\times (2\pi /a)$.
The $\sigma_1$ band is described by the Hamiltonian (\ref{hexa}) with $v=1.996a/\hbar$ and $\gamma =0$.

Third, the $\pi$ band also possesses a circular shaped Fermi surface with $k_5=k_6=0.375\times (2\pi /a)$.
Around  the $\Gamma$ point,  the Hamiltonian can be reduced to 
\begin{equation}
H=(3-(ak)^2)V_{pp\pi}\sigma_x+\varepsilon_p.
\end{equation}
We get the following energy for the above Hamiltonian,\\ 
$E=\varepsilon_p\pm (3-(ak)^2/4)V_{pp\pi}$,
leading to a circularly shaped Fermi surface with the radius 
$k=\sqrt{3+\varepsilon_p/V_{pp\pi}}/a=0.355\times (2\pi /a)$. 
It is in good agreement with  $k=0.375\times (2\pi /a)$ obtained from the first-principles calculations.

 \begin{figure}[t!]
 \begin{center}
 \includegraphics[width=0.5\textwidth]{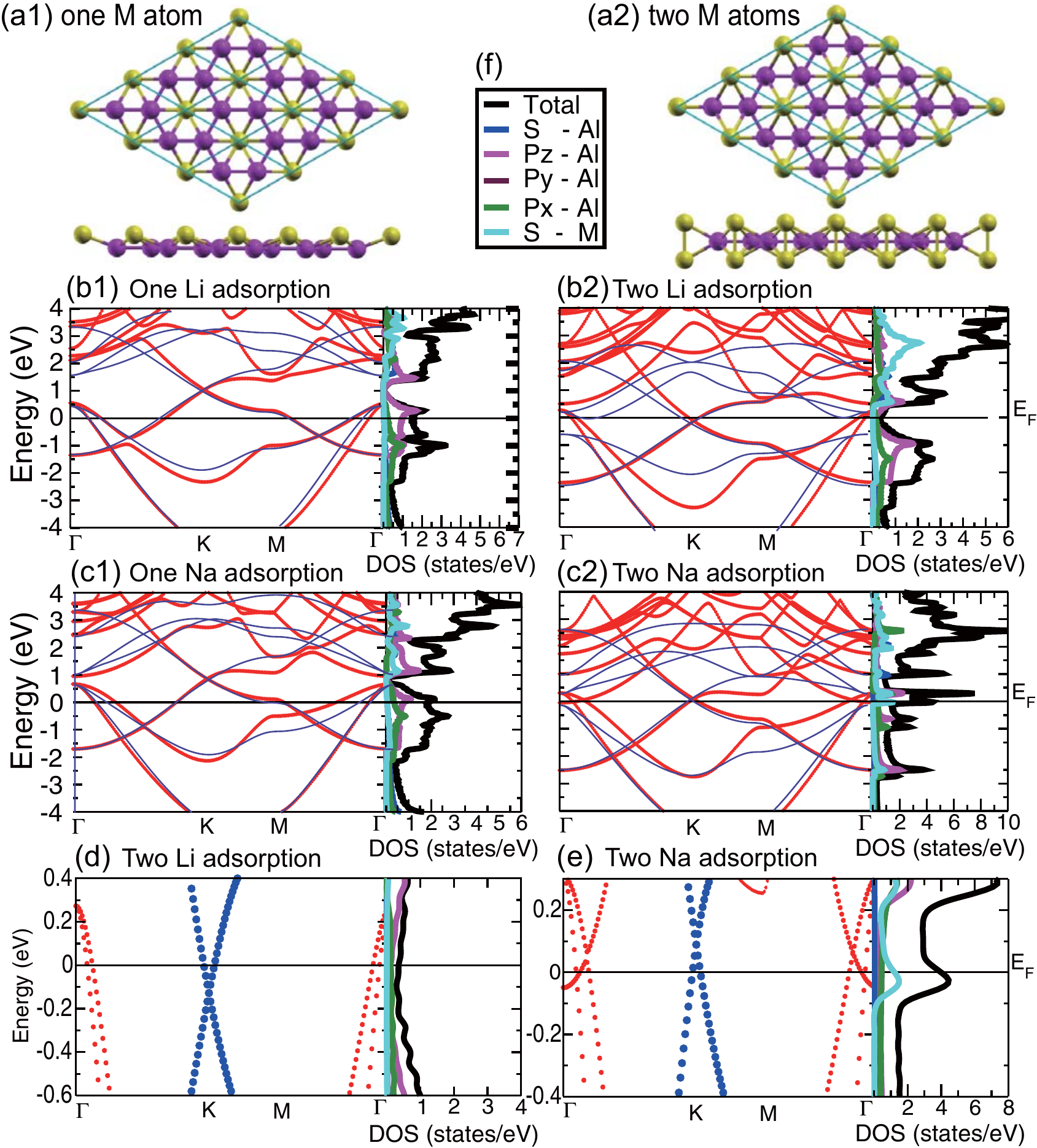}
 \end{center}
 \caption{Aluminene adsorbed with alkali metal atom (M = Li and Na). Geometry of (a1) single and (a2) two M atoms kept at top of the hexagon. 
Electronic band structures of (b1) single and (b2) for two Li adsorption, and  (c1) single and (c2) for two Na adsorption for the left panel, and the corresponding total and partial DOS for the right panel.
 Red curves are obtained from the first-principles calculation.
 Thin black curves are obtained from the tight-binding model, presenting good agreement for one M adsorption but not for two M adsorption. Enlarged version of bands and DOS for (d) two Li atoms and (e) two Na atoms adsorbed on aluminene near the Fermi level.
(f) Color panel for DOS.  }
 \end{figure}

\textbf{Adsorption with alkali-metal atoms:} In the preceding discussions, we have observed that aluminene behaves-like highly hole doped graphene and the Dirac point lies above the Fermi level. Then, it is expected that electron doping in aluminene may shift the Fermi level near to the Dirac point. For this purpose, we adsorb alkali-metal atoms (M = Li and Na) on aluminene and then investigate its effects on the electronic properties of aluminene. Alkali-metal atoms have a tendency to give its valence electron due its free-electron-like behavior and thus they can act as an electron dopant. We consider three possible locations in planar aluminene, namely hollow (above hexagon), top (above Al atom) and bond (above Al-Al bond) for doping alkali metal atoms. We find that the hollow position gives the minimum energy configuration. Geometric structures of aluminene with one and two dopant atoms in hollow positions are shown in Figs.4(a1) and (a2). Results of their geometric analysis and cohesive energies are summarized in Table I.

We show the electronic band structures and DOS obtained from the first-principles calculation in Figs.4(b1) and 4(b2) for aluminene with single and two Li atoms, respectively. Similar results for aluminene with Na adsorption are given in Figs.4(c1) and (c2).  We find that the Fermi level is getting shifted due to the adsorption of single Li and Na atoms. However, the amount of electron doping due to single atom adsorption is not sufficient to shift the Fermi level close to the Dirac point. We also observe an interesting feature that the van-Hove singularity exists at $0.116$ eV for one Na adsorption, which is very close to the Fermi energy. As a result of adsorption two alkali atoms, the Fermi level has shifted close to the Dirac point as shown in Figs.4(d) and (e).

We also fix the Slater-Koster parameters for alkali-metal adsorbed aluminene by fitting at the $\Gamma$ and $K$ points (see Table II). The fitting is reasonably good for one atom doping, as shown in Figs.4(b1) and (c1). However, the fitting at the Fermi energy in the vicinity of the $\Gamma$ point is not so good for two atoms doping, as shown in Figs.4(b2) and (c2). We observe that the bands in the vicinity of the $K$ point move downwards due to alkali-metal atom doping. Consequently, $\varepsilon_p$ changes drastically due to the adsorption of alkali-metal atoms, while there are slight changes in the other parameters (see Table II). As a result, the adatom of alkali-metal atoms acts as the electron dopant. There are only slight changes in the bands near the $\Gamma$ point and they imply that the character of $\sigma$ bonds are not affected by chemical doping. 

Adding one alkali-metal atom per unit cell is not enough to make $\varepsilon_p=0$, since at the maximum it can give $1/2$ electron to each aluminium atom. On the other hand, the Dirac cone exists at the $K$ point with $\varepsilon_p \approx 0$ when we adsorb two alkali-metal atoms per unit cell since they can give $1$ electron to each aluminium atom. Unfortunately, even after the electron doping from the alkali metal atoms, the two sigma bands at the $\Gamma$ point are not completely filled which hinder the Dirac nature emerging in the system. Thus, alkali metal atom adsorbed aluminene behaves as metal unlike the semi-metallic character of graphene.  

\textbf{Conclusion:}
We have demonstrated the stability of graphene-like honeycomb structure made up of aluminium  - planar aluminene - from DFT based phonon and cohesive energy calculations. It is a metal with partially filled $\sigma$ and $\pi$ bands. Our band structure calculations indicate that aluminene can be considered as a highly doped graphene. Fermi surface of this system contains features which correspond to free-electron-like and hexagonal warped surfaces. We obtain Slater-Koster parameters from the tight-binding model and construct the Dirac theory to explain the hexagonal warped Fermi surface of aluminene.  Our investigation suggests that the Fermi energy can be shifted due to strong electron doping from the alkali-metal atoms.

In passing we point out a remarkable feature of aluminene with respect to van-Hove singularities.
They emerge near the Fermi energy. They exist at $E=0.856$eV in the $\pi$ band and at $E=-0.212$ eV in the $\sigma$ band. 
In the case of one Na atom adsorption the van-Hove singularity of the $\sigma$ band emerges at $0.116$ eV. It is intriguing to note that chiral superconductivity is predicted to occur at the van-Hove singularity of the $\pi$ band in graphene\cite{Chubukov}. We might expect a similar phenomenon to occur in aluminene. 
However, a study of this intriguing features is beyond the scope of the present work.

\bigskip

C.K. and A.C. thank Dr. G.S. Lodha and Dr. P.D. Gupta for support and encouragement. 
They also thank the Scientific Computing Group, RRCAT for their support. 
M.E. is very much grateful to Prof. N. Nagaosa for many helpful discussions on the subject.
He thanks the support by the Grants-in-Aid for MEXT KAKENHI Grant Number 25400317.


\end{document}